\documentclass[referee]{raa}            

\usepackage{graphicx,times}             
\usepackage{natbib}
\usepackage{amssymb,amsmath}
\bibpunct{(}{)}{;}{a}{}{,}

\usepackage[a4paper=true]{hyperref}
\hypersetup{colorlinks = true, linkcolor = green, anchorcolor = red, citecolor = blue, filecolor = red, pagecolor = red, urlcolor = red}

\begin{document}

   \title{A triplet of the only pulsation mode detected in the DAV star G132-12}

   \volnopage{Vol.0 (20xx) No.0, 000--000}      
   \setcounter{page}{1}          

   \author{Wenchao Su
      \inst{1}
   \and Jianning Fu
      \inst{1}
   \and Jianxing Chen
      \inst{2,3}
   \and Lester Fox-Machado
      \inst{4}
   \and Shijie Zhao
      \inst{1}
   \and Carmen Ayala-Loera
      \inst{4,5}
   \and Jiangtao Wang
      \inst{1}
   \and Yang Pan
      \inst{1}
}

   \institute{Department of Astronomy, Beijing Normal University,
             Beijing 100012, China; {\it jnfu@bnu.edu.cn}\\
        \and
             Dipartimento di Fisica e Astronomia, Alma Mater Studiorum Universita' di Bologna, via Piero Gobetti 93/2, I-40129 Bologna, Italy\\
        \and
             INAF-Osservatorio di Astrofisica e Scienze dello Spazio di Bologna, Via Piero Gobetti 93/3 I-40129 Bologna, Italy\\
        \and
             Instituto de Astronomía, Universidad Nacional Autónoma de México,  Ensenada, BC, México\\
        \and
           Institute of Astronomy \& Astrophysics Academia Sinica, Taipei\\
\vs\no
   {\small Received~~20xx month day; accepted~~20xx~~month day}}

\abstract{ 
Hydrogen atmosphere pulsating white dwarfs, also known as DAV stars, are the most abundant type of pulsating white dwarfs.
High-temperature DAV stars exhibit in general a small number of pulsation modes and stable frequencies.
G132-12 is one of the pulsating hydrogen atmosphere white dwarf stars which lies close to the blue edge of the instability strip.
Previous researches reported that G132-12 might have only one pulsation mode with the period of 212.69 s.
To study the pulsation properties of G132-12 in detail, we carried out a bi-site observation campaign in October 2019.
Time series photometric data were collected during around 154 hours in total.
A Fourier Analysis reveals 3 frequencies which are identified as the triplet of a $l = 1$ g-mode pulsation with the period of 212.499 s.
The rotational period is derived as $P_{rot} = 35.0\pm6.7$ hours and the inclination of the rotational axis to the line of sight is $70^{\circ}$. 
G132-12 could be an ideal target for measuring the cooling scale of this white dwarf star with only one excited pulsation mode detected.
}

   \authorrunning{Wenchao Su et al.}            
   \titlerunning{Pulsations of G132-12}  

   \maketitle

\section{Introduction}           
\label{sect:intro}

White dwarfs are the final products of the majority of stars in the universe. About 95\% stars should end their lives as C/O-core white dwarfs.
The white dwarfs with hydrogen atmosphere, also named DA type white dwarfs, which account for about 74\% of the number of known members \citep{2019A&ARv..27....7C}, are the most classical type of white dwarfs.

Since the discovery of the first pulsating hydrogen atmosphere white dwarf star (DAV or ZZ Ceti star) HL Tau 76 \citep{1968ApJ...153..151L}, 270 DAV stars have been confirmed \citep{2019A&ARv..27....7C,2020IAUS..357..123V}.
These stars define an experimental instability strip in the H-R diagram \citep{2015ApJ...809..148T} with temperatures ranging in 10500 K - 13000 K \citep{2019MNRAS.482.4018B}.
They show different pulsation characteristics on different positions in the instability strip, from the blue edge to the red edge:
i) DAV stars close to the blue edge of the instability strip show a few modes with periods of 100–300 s and low amplitudes of about 1 mmag,
ii) For the DAV stars cooler by a few hundreds of degrees than the members close to the blue edge, they show short-period pulsations with larger amplitudes,
iii) DAV stars located in the middle of the instability strip show pulsations with periods up to 300 s and even larger amplitudes.
There are usually a large number of nonlinear combination frequencies detected in the amplitude spectra,
iv) The cooler DAV stars have longer but more unstable pulsation periods. Meanwhile, outbursts start to be detected,
v) The coolest DAV stars tend to have the longest pulsation periods with relatively low amplitudes ($<$ 1 mmag)\citep{2017ApJS..232...23H,2019NewA...7301276C}.

G132-12, also named EGGR167, was discovered to be a DAV star based on 10 hours of time-series photometry by \cite{2006AJ....132..831G} with only one period of 212.7 s detected.
As far as the atmosphere parameters of G132-12, \cite{2006AJ....132..831G} reported the effective temperature of $12080\pm145$ K and log $g$ of $7.94\pm 0.04$.
After that, \cite{2011ApJ...743..138G} provided a new result for G132-12 through a new generation 3-D atmosphere model with the effective temperature of $12610\pm222$ K and log $g$ of $8.01\pm0.06$.
\cite{2019NewA...7301276C} collected nearly 90 hours of time-series photometry from single-site observations in 2017 and detected also only one period of 212.69 s.

For the DAV stars close to the blue edge, the pulsations show few stable modes. The high-temperature DAV stars such as G117-B12A and R548 have been proven to be stable optical clocks, with stability timescales of the order of 2 Gyr \citep{2013ApJ...771...17M,2021ApJ...906....7K}.
G132-12, as a DAV close to the blue edge, becomes one of interesting targets of our research.
In order to explore the pulsation characteristics of G132-12 in detail, we conducted the first international bi-site observation campaign for G132-12 in 2019.

In this paper, we describe the observations for G132-12 and data reduction in Section \ref{sect:Obs}; In Section \ref{sect:analysis}, we present the results of Fourier analysis; Finally, we make a discussion in Section \ref{sect:discussion} and present a summary in Section \ref{sect:summary}.

\section{Observations and data reduction}
\label{sect:Obs}

In October of 2019, we carried out time-series photometric observations for G132-12 with both the 85-cm telescope at Xinglong Station (XL) of National Astronomical Observatories of China and the 84-cm telescope at San Pedro Martir (SPM) Observatory of Mexico.
Johnson V filters were used and the 40 s of exposure time was applied for the bi-site observations.
In total, about 154 hours of data were collected. The journal of observations is listed in Table \ref{tab:t1}.

\begin{table}
    \centering
	\caption{Journal of the bi-site photometric observations of G132-12 \label{tab:t1}}
	\begin{tabular}{c|cc}
	\hline
	\hline
	UT\_Date & \multicolumn{2}{c}{Observation-Duration}\\
	 & SPM-84cm & XL-85cm \\
    (October 2019)& (Hours) & (Hours)\\
	\hline
	19    & 8.71&\\
	20    & 9.63&  9.71 \\
    21    & 9.51&  9.67 \\
    22    & 9.43&  8.50 \\
    23    & 8.68&  4.15 \\
    24    & 8.76&  6.45 \\
    25    & 9.10&  9.33 \\
    26    & 9.20&  9.17 \\
    27    & 9.13&  5.81 \\
    28    &     &  8.87  \\
    \hline
    Total & 82.15& 71.66\\
    \hline
    \hline
    \end{tabular}
\end{table}

\begin{figure}
	
	\begin{center}
		
		\includegraphics[width=0.7\textwidth]{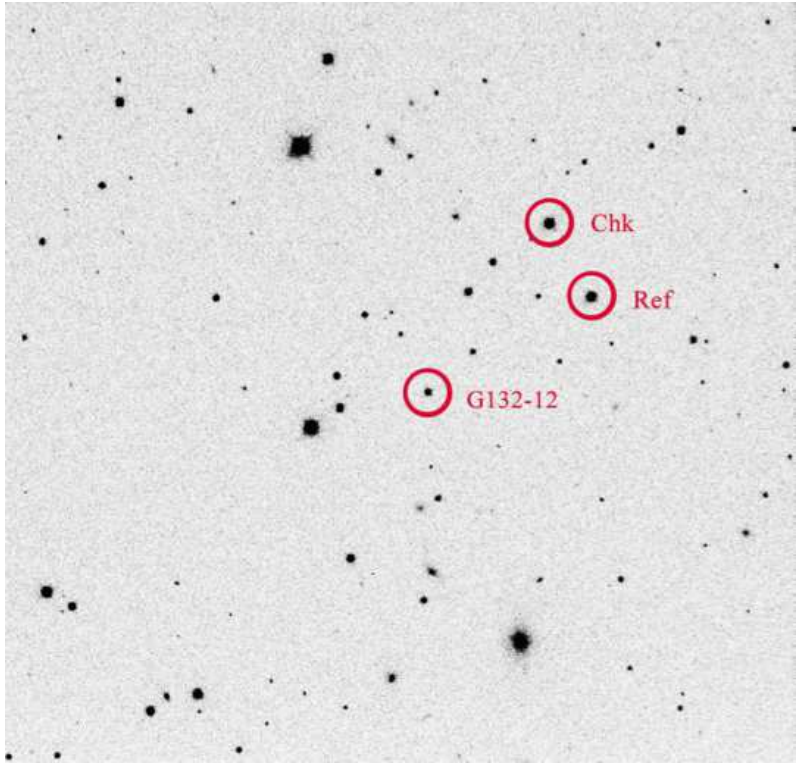}
		
	\end{center}
	
	\caption{An original observation image with the SPM-84 cm telescope. The target star G132-12, the reference star and the check star are marked with red circles.}
	\label{fig:t1}
	
\end{figure}

We perform data reduction of all the photometric data with IRAF (Image Reduction and Analysis Facility) package \citep{1986SPIE..627..733T}.
First, the pre-reduction for bias and flat field is made to the original images.
As the exposure time is shorter than 1 minute, the dark current is ignored.
We perform aperture photometry on the science images with APPHOT task under IRAF package. 
We select two constant stars as reference star and check star in the field, respectively, which is the same selection as \cite{2019NewA...7301276C}, as shown in Figure \ref{fig:t1}. 

\begin{figure}
     \begin{center}
	\includegraphics[width=0.7\textwidth]{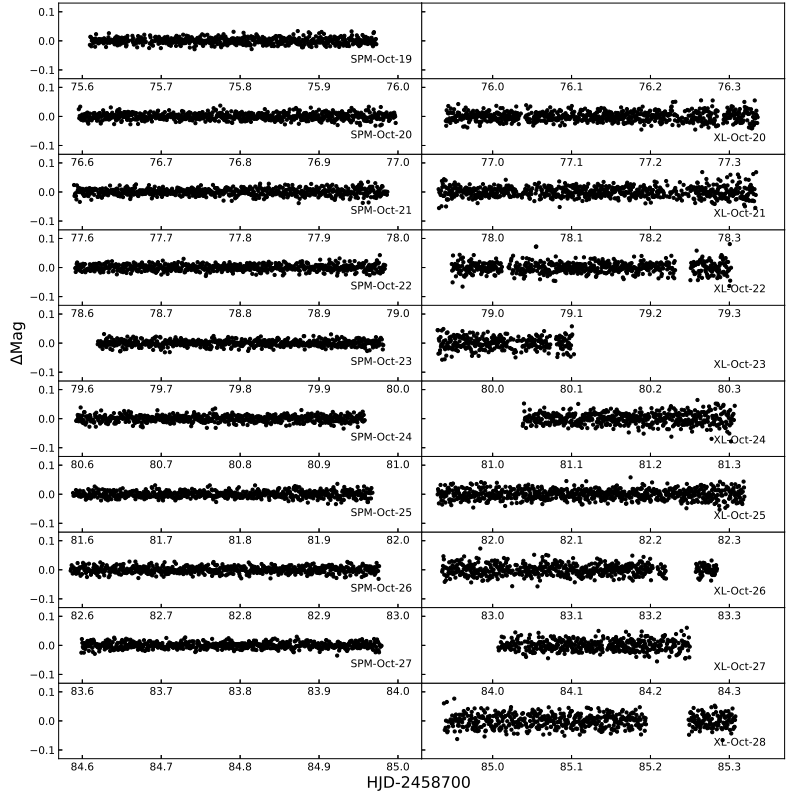}
     \end{center}
	\caption{Light curves of G132-12 from the SPM-84 cm telescope and the XL-85cm telescope in October 2019. The abscissa is HJD (Heliocentric Julian Day) - 2458700.  }\label{fig:t2}
\end{figure}

In order to determine the optimal aperture, we measure each image with a step of 0.1 times of FWHM between 0.5 and 3 times to obtain different time-series of differential magnitudes between the two constant stars.
We calculate the standard deviations of these time series separately and select the aperture corresponding to the smallest standard deviation as the optimal aperture.
Considering the data obtained from bi-site observations, we subtract the average value of each night's light curve to avoid the difference of zero points caused by different facilities in different nights.
In addition, as changes in atmospheric transparency and instability of the instrument system during overnight observations may cause long-term trends in the light curves, we make calibration by dividing the light curves by the fitted polynomials.
Considering the difference of observation conditions in different nights, we use two or three order polynomial fitting in order to address long-term trends for each light curve.
The reduced light curves from SPM and XL observatories are shown in Figure \ref{fig:t2}.

\section{Amplitude spectra and frequency analysis}
\label{sect:analysis}

\begin{figure}
	\includegraphics[width=0.6\textwidth]{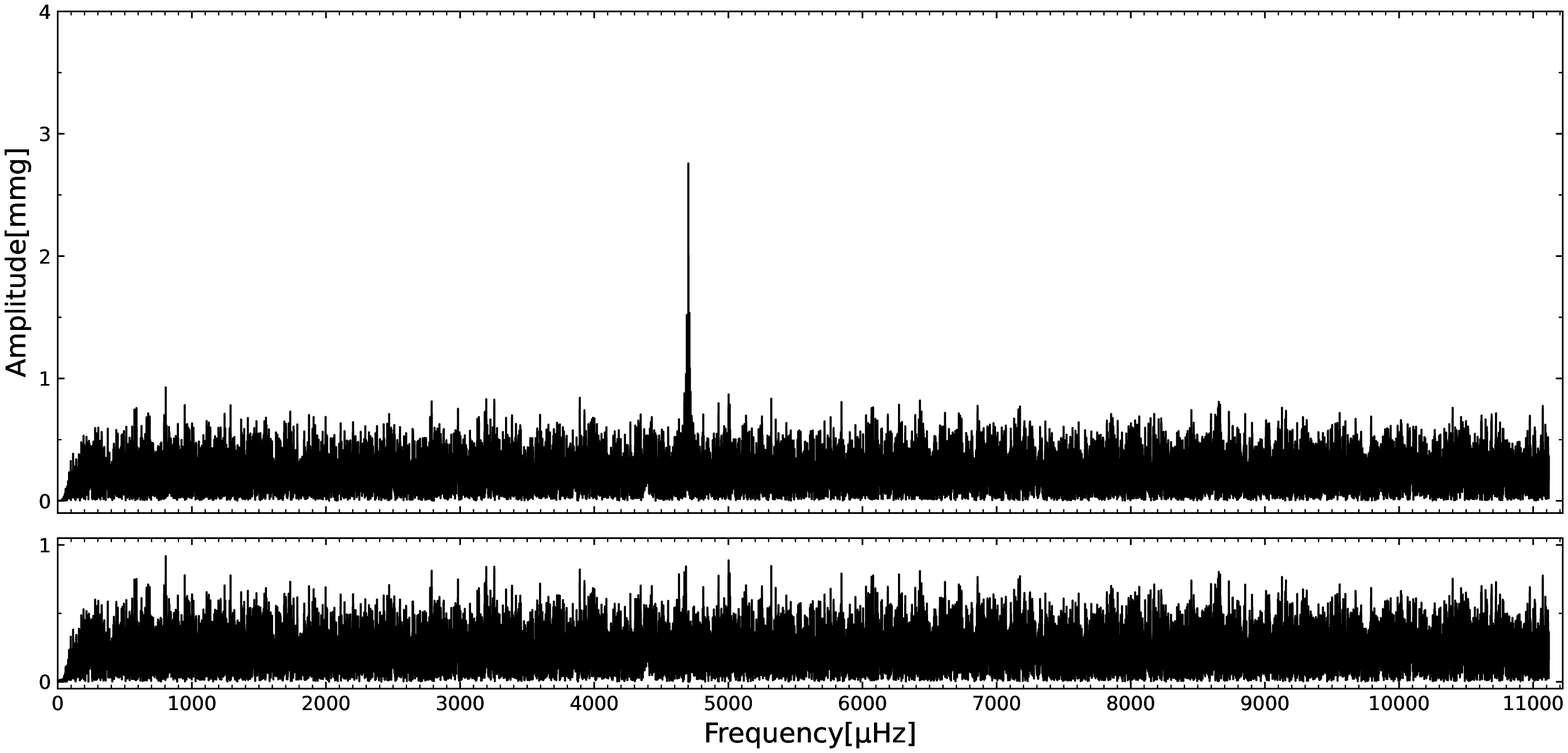}
	\includegraphics[width=0.4\textwidth]{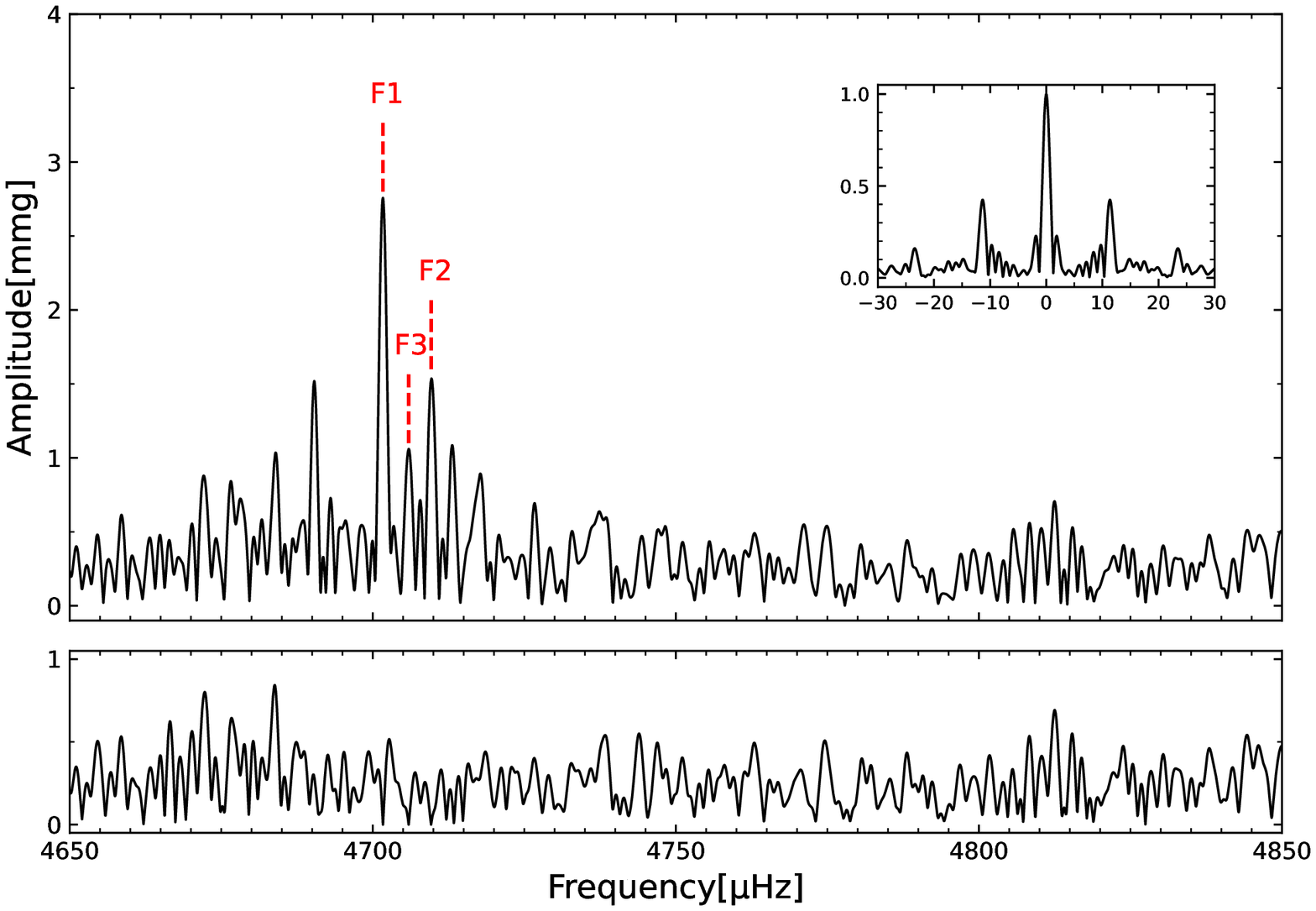}
	\caption{Left: Amplitude spectra of G132-12 from light curves with bi-site data in Oct 2019. Right: Enlarged view of the part of the spectrum near the significant peaks. The spectral window is shown as the inset on the right top. Three frequencies are marked in red color.}
	\label{fig:t3}
\end{figure}

We perform Fourier transformation on the total light curves of G132-12 with the Period04 program \citep{2005CoAst.146...53L}.
Bi-site observations suppress effectively the influence of daily aliasing.
According to the acceptance standard with a signal-to-noise ratio larger than 4 \citep{1993A&A...271..482B,1997A&A...328..544K}, we detect in total three frequencies from the Fourier spectrum: 4701.68 $\pm$ 0.17 $\mu$Hz, 4705.91 $\pm$ 0.47 $\mu$Hz, and 4709.64 $\pm$ 0.33 $\mu$Hz.
The Fourier spectrum and the spectrum of the residuals after the three frequencies are prewhitenned are shown in Figure \ref{fig:t3}.
The detected frequencies are listed in Table \ref{tab:table2}.
The uncertainties of frequencies and amplitudes are calculated with a Monte-Carlo simulation \citep{2013MNRAS.429.1585F}.

As the frequency intervals between the three frequencies of G132-12 are approximately equal and the average is 3.98 $\pm$ 0.65 $\mu$Hz, we identify that the three frequencies as a triplet due to rotation with $l=1$.
The frequency of the eigenmode should be 4705.91 $\pm$ 0.33 $\mu$Hz corresponding to the period of 212.499 s.
The mode with the period of 212.7 s detected by \cite{2006AJ....132..831G} and the one with the period of 212.69 s detected by \cite{2019NewA...7301276C} should be the split mode with $m = -1$.

\begin{table}
    \centering
	\caption{Frequency list of G132-12}
	\label{tab:table2}
	\begin{tabular}{ccccc}
	    \hline
	    \hline
	    ID & Frequency & Amplitude & Period & SNR\\
	     & ($\mu$Hz) & (mmag) & (s) & \\
        \hline
	    F1 & 4701.68 $\pm$ 0.17 & 2.7 $\pm$ 0.13 & 212.690 $\pm$ 0.007 & 10.64\\
        F2 & 4709.64 $\pm$ 0.33 & 1.5 $\pm$ 0.14 & 212.330 $\pm$ 0.015 & 6.05\\
	    F3 & 4705.91 $\pm$ 0.47 & 1.0 $\pm$ 0.13 & 212.499 $\pm$ 0.021 & 4.05\\
	    \hline
	\end{tabular}

\end{table}

For g-mode non-radial pulsations, if all terms higher than the second order are negligible, frequency splitting can be expressed by the following formula \citep{1992ApJ...394..670D}:
\begin{equation}
\omega_{k,l,m} = \omega_{k,l} - m\Omega(1-\frac{1}{\eta}).
\end{equation}
where $\omega$ is the frequencies of different $m$-values of the mode under the same $l$-value and $k$-value; $\eta = l (l + 1)$, $\Omega$ is the solid-body rotation angular velocity of the star.

From the calculation, the rotation period of G132-12 is derived as 35.0$\pm$6.7 hours.
This is in agreement with the mean value of rotational periods from asteroseismology studies of a 31 known pulsating white dwarfs \citep{2017ApJS..232...23H}. 
In the 31 samples summarized by them, the mean rotation period is about 35 hours.

In addition, assuming that the ratio of amplitude of a triplet is only due to geometric effects, and the pulsating axis is aligned with the rotation axis of this star, we can estimate the angle of the rotation axis and sight line \citep{2013MNRAS.429.1585F}.
In this triplet, the central mode ($m=0$) shows a weaker amplitude than the $m=-1$ and $+1$ modes.
Then we calculate the ratio of the average amplitudes of the $m=\pm1$ mode to that of the $m=0$ mode.
The result is $2.1$, which means that the angle of the rotation axis to the sight line is about $70^{\circ}$ \citep{1985ApJ...292..238P}.

\section{Discussion}
\label{sect:discussion}

\subsection{\textit{TESS} data}
The Transiting Exoplanet Survey Satellite (\textit{TESS}), an all-sky survey mission, records pulsation signatures from some bright white dwarfs over the entire sky.
We noted that G132-12 (TIC 267687713) has been observed in sector 17 by \textit{TESS} at 2-minute cadence.
We downloaded all 2-minute light curves of this star for asteroseismic analysis.
Unfortunately, there is not any obvious signal in the Fourier spectrum.
Considering that only 10 cm effective pupil diameter of \textit{TESS}, we believe that this is because the luminosity of G132-12 is too low ($M_{\textit TESS}=17.33$ $mag$) for \textit{TESS} to observe the pulsations effectively.
In addition, the 2-minute observation cadence of \textit{TESS} exceeds half of the pulsation period of G132-12, it also brings the difficulty to detect the pulsations.

\subsection{Measuring the cooling rate}

Non-radial g-modes of high-temperature DAV stars often show extremely slow period changes caused by the gradual cooling of stars, which can be used to measure the cooling rate of white dwarf stars.
G117-B15A and R548 (ZZ Ceti itself) are two of the long-studied high-temperature DAV stars which have been observed since 1970s.
Six independent modes had been detected in G117-B15A.
The rate of period change, $dP/dt$, of the 215.2 s mode in G117-B15A was measured to be $(5.12\pm0.82)\times10^{-15} s\cdot s^{-1}$ with the $O-C$ method \citep{2021ApJ...906....7K}.
For R548, \cite{2013ApJ...771...17M} found 5 independent modes and measured $dP/dt$ of the 213.1 s mode as $(3.3\pm1.1)\times10^{-15} s\cdot s^{-1}$.
The two stars show considerable stable frequencies in their largest-amplitude modes. 
Furthermore, \cite{2013ApJ...766...42H} reported a rapidly cooling DAV star WD 0111+0018. After more than nine years of monitoring, they measured $dP/dt$ for each of the two independent modes.
All periods are changing at rates faster than $10^{-12} s\cdot s^{-1}$. 

Constructing the $O-C$ diagram is a common method for measuring the long-term period changes of DAV stars. If we assume that all terms higher than the second order are negligible, we get the $O-C$ equation following \cite{1991ApJ...378L..45K}:\begin{equation}
O-C=\Delta t_{0}+\Delta P_{0}E+\frac{1}{2}P_{0}\dot P E^{2}.
\end{equation}
where $\Delta t_{0}$ is the uncertainty in the first maximum, $P_{0}$ the period at the first maximum, $\Delta P_{0}$ the error in the period $P_{0}$. 

It should be noted that for multi-mode stars, the stable largest-amplitude modes are selected to measure $\dot P$ in general,  since the maximum times $O$ of the largest-amplitude mode at different epoch $E$ can be obtained from the observed light curves.
Because light curves are coupled by all its independent modes, the change of any mode will affect the result of the Fourier transform of light curves.
The calculations of $O-C$ caused by the drift of other modes have inherent errors that cannot be eliminated.
Hence, if confirmed as a one-mode DAV star by long-term observations in the following years, G132-12 could be an ideal object for measuring the period change rate and the cooling scale of white dwarf stars.

\section{summary}
\label{sect:summary}
We obtain high-cadence photometric data for the DAV star G132-12 through 154 hours of time-series bi-site observations.
After data reduction and Fourier analysis, we detect three frequencies with the SNR larger than 4 and identify them as a triplet due to rotation.
This is for the first time that the rotation of G132-12 is detected.
We calculate the rotation period of 35.0 $\pm$ 6.7 hours. 
Furthermore, we correct the period value of the only mode detected as 212.499 $\pm$ 0.021 s.
In addition, we estimate the angle of the rotation axis to the line of sight at about $70^{\circ}$.

G132-12 is worthy of long-term observations and deep research.
Comparing the pulsation period in 2006, 2017, and 2019, the only mode of G132-12 showed considerable stability, which means that it could be the ideal optical clock whose rate of period changes can be accurately measured.
Moreover, we hope that there will be more photometric observations with larger ground-based telescopes in following years for this star to confirm G132-12 is the first one-mode DAV pulsator. 
If confirmed, G132-12 is promising to define the blue edge of the instability strip of DAV stars.
The pure frequency spectrum with only one mode will allow one to get an excellent measurement of the cooling time scale without inherent errors of multiple modes, which may help to limit the lower boundary of the age of the Galaxy and the universe.

\begin{acknowledgements}
JNF acknowledges the support from the National Natural Science Foundation of China (NSFC) through the grants 11833002,12090040 and 12090042.
LFM and RM acknowledge the financial support of the UNAM under grant PAPIIT IN100918.

WS would like to thank Dr. LFM for his hospitality during research stay in Mexico at Instituto de Astronom{\i}a de la UNAM, Campus Ensenada.

\end{acknowledgements}

\label{lastpage}

\end{document}